\documentstyle[aps,prl,twocolumn,epsf,epsfig]{revtex}
\def\beq{\begin{equation}}
\def\eeq{\end{equation}}
\def\beqarr{\begin{eqnarray}}
\def\eeqarr{\end{eqnarray}}

\begin{document}
\draft
\twocolumn[\hsize\textwidth\columnwidth\hsize\csname @twocolumnfalse\endcsname

\title{Vortex States of a Superconducting Film from a Magnetic Dot Array} 
\author{D. J. Priour Jr$^{1}$, H. A. Fertig$^{2}$}
\address{$^1$Condensed Matter Theory Center, Department of Physics,
University of Maryland, College Park, MD 20742-4111\\
         $2$Department of Physics and Astronomy, University of Kentucky,
Lexington, KY 40506-0055}

%\date{\today}

\maketitle

\begin{abstract}
Using Ginzburg-Landau theory, we find novel configurations of vortices
in superconducting thin films subject to the magnetic field of a magnetic
dot array, with dipole moments oriented perpendicular to the film.
Sufficiently strong magnets cause the formation of vortex-antivortex pairs.
In most cases, the vortices are confined to dot regions, while the 
antivortices can form a rich variety of lattice states.  We propose an
experiment in which the perpendicular component of the dot dipole moments
can be tuned using an in-plane magnetic field.  We show that in such an 
experiment the vortex-antivortex pair density shows broad plateaus as a 
function of the dipole strength.  Many of the plateaus correspond to 
vortex configurations which break dot lattice symmetries.  In some of 
these states, the vortex cores are strongly distorted.  Possible 
experimental consequences are mentioned.
\end{abstract}

%\pacs{}
%\vspace{-.5cm}
]

%{\it Introduction--}
    Type II superconductors, with their high critical currents and 
fields, lend themselves to technological applications.  However, vortices
appear in these superconducting systems when magnetic fields or currents
are made sufficiently large.  Motion of the vortices spoils the perfect
conductivity important in applications; it is therefore important to 
find ways to pin flux quanta.  Systematic studies of vortex pinning 
have been carried out in experiments on regular nanoscale arrays where
a lattice of defects is superimposed on a thin superconducting film.
``Antidot'' arrays, in which pinning centers consist of holes or 
depressions in the substrate, were first to be studied~\cite{baert}.  
Subsequently,
magnetic dot arrays have been created, usually by the deposition of 
mesoscopic magnetic dipoles on top of the superconducting 
film~\cite{schuller,bael,martin,bael2,lange1,mosh}.  In the case of the 
nanoscale magnetic dot arrays, 
a number of experimental and theoretical studies
have focused on scenarios in which each unit cell is penetrated by 
a finite amount of net magnetic flux.  In this work, we consider
arrays of dipoles for which there is no net perpendicular applied field.  Even
without any applied net flux, we show that rich vortex phenomena occur
as one changes the strength of the dot dipoles.  We propose an 
experimental scenario to observe these effects, in which an in-plane
applied magnetic field is used to tilt the dipole moments.  The 
thin film geometry prevents the horizontally directed field from disturbing the 
superconducting state, and tilting the moments makes possible the adjustment 
of the effective strength of the magnetic dots. 

    In our work, we focus on the case in which the dipoles 
prefer an orientation perpendicular to the superconducting film.  The 
supercurrents which the magnetic dots induce move in a clockwise 
direction.  The resulting cancellation of the field from the array 
magnets is a partial realization of the Meissner effect.  With the
induced currents, there is an associated cost in kinetic energy; hence, 
for sufficiently strong dipoles, it is energetically favorable to 
have a vortex in the vicinity of the dot, since this allows the
vortex's counterclockwise currents to partially cancel the induced 
currents.  However, due to the zero flux condition, vortices cannot
form in isolation; vortices and antivortices must be generated in 
pairs.

     Some work has concentrated on isolated magnets, where the entire
system has cylindrical symmetry~\cite{marmorkos,Peeters1,Peeters2}.  A few 
studies in the framework of the London theory have focused on a single
pair of flux quanta in one unit cell, using periodic boundary 
conditions~\cite{Cheng,hwa,erdin1,erdin2,Erdin,Carneiro}.  
Depinning has been studied 
in the Ginzburg-Landau framework, but for antidots rather than magnetic 
dots~\cite{priour}.  Because vortices must be put in by hand in the 
London theory, is it not straightforward to go beyond simple situations
(generally one vortex-antivortex pair in a single unit cell).  However
vortices appear naturally in a Ginzburg-Landau treatment, making tractable
more complicated situations, such as those involving multiple flux quanta 
pairs in a unit cell, where often one must allow for superlattice structures.
In this work, to avoid assuming the same vortex configuration for each 
unit cell, we study a large ($4 \times 4$) supercell with periodic 
boundary conditions.  In what follows, $\rho_{pair}$ denotes the 
number of vortex-antivortex pairs per unit cell.  We find that as the strength
of the dot magnets is varied (by tilting dipole moments via an
in-plane magnetic field), $\rho_{pair}$ exhibits sharply defined plateaus.
Remarkably, the vortex pair density is not always a monotonic function of 
the dipole strength.
We will see that there are stable vortex states which break orientational
and/or mirror symmetry.  Phase transitions which can be abrupt or continuous
occur as the dipole strength is varied.  In the abrupt case, hysteresis 
phenomena are found, consistent with their being first order transitions.  
Surprisingly, some of these occur within a plateau.  As will be discussed,
second order transitions define shifts between plateaus in which $\rho_{pair}$
either increases or decreases via the annihilation or creation of a 
vortex-antivortex pair.  A gradual deformation of the vortex and antivortex
cores is associated with the creation or destruction of a flux quantum pair.
Since London theory does not take into account variations of the Cooper
pair density, and therefore cannot correctly describe vortex cores, 
Ginzburg-Landau theory is essential to describe these novel states.  

{\it Methods and Results--}
      To study our system in the Ginzburg-Landau framework, we solve the 
nonlinear partial differential equations which one finds on extremizing 
the Ginzburg-Landau internal energy given by
\begin{equation}
E_{GL} = d \Delta \xi^{3} \int \left[ \begin{array}{c} \left| \psi^{*} \left(
\vec{\nabla}/i - \vec{A} \right) \psi \right|^{2} - \left| \psi \right|^{2}
\\ + \frac{\kappa^{2}}{2} \left| \psi \right|^{4} + (\vec{B}-\vec{B}_{ext})^{2}
\end{array} \right] d^{3}x.
\label{Eq:eq1}
\end{equation}
In Eq.~\ref{Eq:eq1}, dimensionless units are used; as a result, all linear 
dimensions are expressed in terms of the superconducting coherence length
$\xi$.  The constant $\Delta$ is the condensation energy per unit volume
for a uniform bulk superconductor and the film thickness (in units of $\xi$)
is given by $d$.  Since $d \ll \xi$, we view the superconducting substrate
as a film of negligable thickness.  Hence, one would not expect an 
in-plane magnetic field imposed to tilt the dipole moments of the dot 
magnets to affect the superconducting state in the film.   
We assume the mesoscopic magnetic dipoles above the substrate
to have a square cross section.  While there is a range of magnet  
thicknesses, we have chosen our dots to be cubic in shape with
$2.0 \xi$ as the length of a side.  In our case, the mesoscopic 
magnetic cubes form a square array whose
lattice constant is $6.25\xi$.  We specify the dipole moment of the magnets 
by calculating the {\it positive} flux passing through each unit cell.  Given 
in units of the fundamental flux quantum, this quantity provides a natural 
measure of the dot dipole strength.  

      In solving the Ginzburg-Landau equations for our geometry, we have 
replaced continuum variables with their discrete versions on a mesh fine 
enough to ensure convergence with respect to the discretization (to 
achieve this, we allow at least 5 grid points per coherence length).  
Our scheme of discretization is a gauge theoretic formalism (discrete 
versions of the standard continuum gauge symmetries are imposed) where currents
and vector potentials $A_{ij}^{x}$ and $A_{ij}^{y}$ occupy the lattice bonds,
while order parameter values $\psi_{ij}$ reside on lattice nodes.  
We treat the mesoscopic magnetic dots as square loops of current with 
a thickness equal to the width of the dot base; the magnetic field and 
vector potential generated by the magnetic dipoles are then easily 
calculated.  Though we handle the $x$ and $y$ (in-plane coordinates) 
discretely, the $z$ direction is treated exactly, in the continuum limit.  

     Via a conjugate-gradient technique, we solve the Ginzburg-Landau 
equations in an iterative manner.  In this method, one first linearizes
the Ginzburg-Landau equations about an initial guess.  The solution obtained
by solving the resulting linear equations is then used as an input for the 
next iteration.  We simulate in our calculation an experiment in which the 
effective magnetic dot strengths are varied continuously (to realize this
in the laboratory, one could as noted above apply an in-plane magnetic field 
to tilt the dipole moments of the dot magnets, thereby varying their effective
strength).  In ``rightward sweeps'', we slowly increase the dipole strength.
Sweeps range from dipoles too weak to generate any vortex-antivortex pairs
to magnetic dots strong enough to destroy the superconductivity altogether.
In a rightward strength (toward stronger dipoles), the results of each 
calculation are used as the initial guess of the next calculation, in which
the Ginzburg-Landau equations are solved for slightly stronger magnetic 
dots.  ``Leftward'' sweeps are conducted in a very similar way, with 
each successive calculation using weaker dipoles than the previous one.
The sweeps in opposite directions complement each other by highlighting 
hysteresis effects, thereby revealing which transitions are first order.

     To illustrate the configurations which correspond to the phases shown 
in Fig.~\ref{Fig:fig2}, we display Cooper pair densities in Fig.~\ref{Fig:fig3}
and Fig.~\ref{Fig:fig4}.  Vortices and antivortices are readily identified
as regions of depleted Cooper pairs.  Exploiting the fact that the order 
parameter phase is singular at the vortex cores makes it possible to 
distinguish vortices from antivortices; the line integral $\oint \vec{\nabla}
\phi \cdot d\vec{s}$ yields $+2\pi$ ($-2\pi$) if a (anti) vortex is surrounded
by the integration contour, and is zero otherwise.  In terms of currents and
magnetic flux, this condition is $\oint \frac{\vec{J}}{\left| \psi \right|^{2}}
\cdot d\vec{s} + \Phi_{B} = \pm 2 \pi$ for a vortex or antivortex, respectively.
This tool for locating flux quanta permits the precise depiction of vortex
configurations, and also conveniently yields $\rho_{pair}$.  

\begin{figure}
\centerline{\psfig{figure=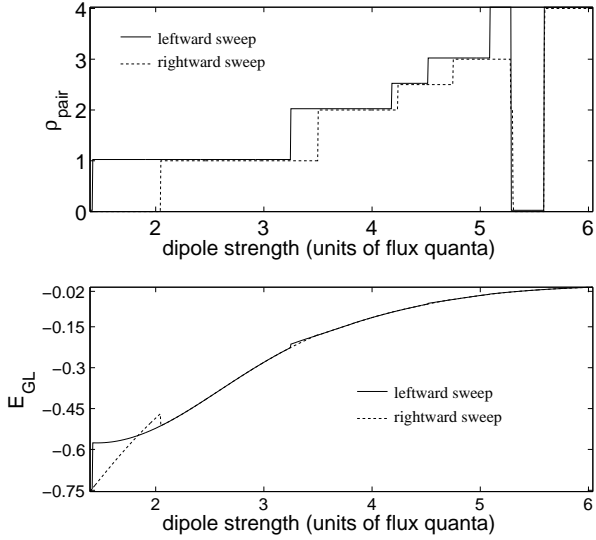,width=3.4in}}
\caption{$\rho_{pair}$ and $E_{GL}$ plots for sweeps to the left and to the 
right.  The energy is given in units of $d \xi^{3} \Delta$, where $\Delta$
is the condensation energy per unit volume and $d$ is the film thickness.}
\label{Fig:fig1}
\end{figure}

\begin{figure}
\centerline{\psfig{figure=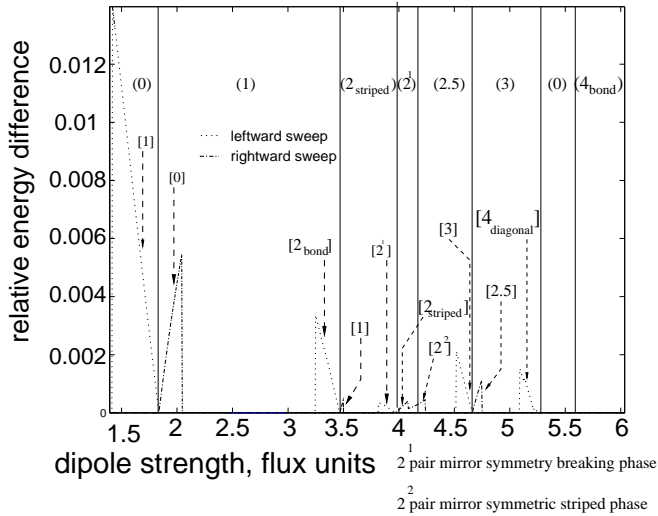,width=3.4in}}
\vspace{1mm}
\caption{Phase diagram with metastable states.  Stable states indicated
by numbers in parenthesis; metastable states by numbers in brackets.
These numbers indicate how many vortex-antivortex pairs per unit cell.
Dotted (broken) lines indicate energies of metastable states for 
leftward (rightward) sweeps.  ``Bond'' subscripts indicate states in which 
antivortices lie along nearest neighbor bonds; ``striped'' and ``diagonal''
subscripts indicate antivortices which lie on diagonals (next-nearest
neighbor bonds).}
\label{Fig:fig2}
\end{figure}

\begin{figure}
\centerline{\psfig{figure=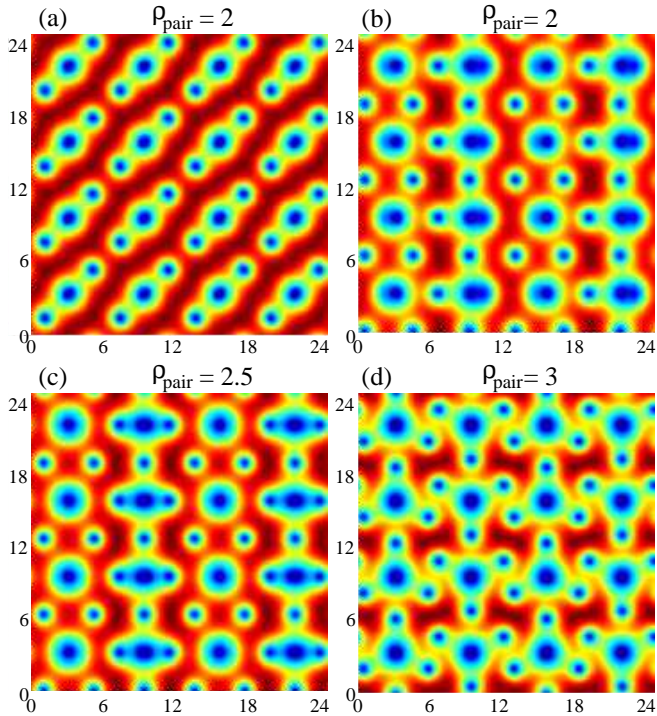,width=3.4in}}
\vspace{1mm}
\caption{Cooper pair density plots of stable phases.  the images in panels
(a), (b), (c), and (d) correspond to dot magnet strengths of $4.10$, 
$4.62$, $4.91$, and $5.62$ fundamental flux units, respectively. 
Antivortices appear as small dark spots, while the large dark spots 
indicate dot regions; Cooper pair density is depressed in regions with
lighter shading.}
\label{Fig:fig3}
\end{figure}

\begin{figure}
\centerline{\psfig{figure=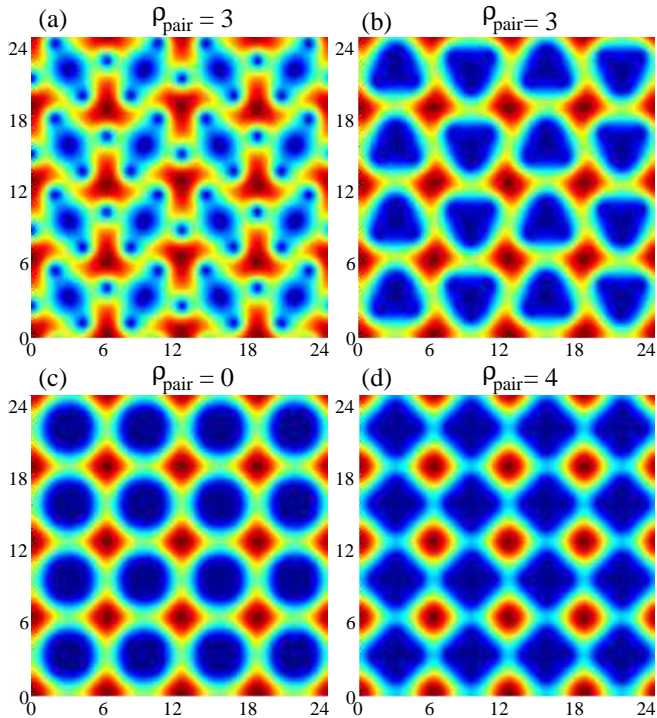,width=3.4in}}
\vspace{1mm}
\caption{Cooper pair density plots of stable phases.  Images in panels
(a), (b), (c), and (d) correspond to magnetic dot strengths equal to 
$5.85$, $5.97$, $6.14$, and $6.85$ fundamental flux units, respectively.}
\label{Fig:fig4}
\end{figure}

     Figure~\ref{Fig:fig1} is a plot of $\rho_{pair}$ and the Ginzburg-Landau
internal energy for both sweeps to the right and to the left ($E_{GL}$ is 
given in units of $d \Delta \xi^{3}$).  The discrepancy between opposite 
sweeps for $\rho_{pair}$ is readily evident.  One can also see sudden 
downward jumps in $E_{GL}$ for both rightward and leftward sweeps.  These
jumps and the hysteresis effects in $\rho_{pair}$ are a result of the 
metastability of some of the vortex configurations, signifying that 
several of the transitions are first order.  One can construct a ``ground
state energy'' by selecting the lowest energy from sweeps to the right and 
to the left, with the preferred state being the vortex configuration 
corresponding to the lowest energy.  In this manner, we have constructed
a phase diagram for the system, shown in Fig~\ref{Fig:fig2}.  
Figures~\ref{Fig:fig3} and~\ref{Fig:fig4} depict vortex configurations in
salient cases.  In the phase diagram, states are classified according to 
$\rho_{pair}$, the number of vortex-antivortex pairs.  When necessary, as
in the case of $\rho_{pair} = 2$, phases are further classified according to 
the symmetry of the vortex configuration.   
metastable states are not lowest in energy and are therefore not preferred by
the system.  
%%%%%%%%%%%%%%%%%%%%%% FIG. 1 %%%%%%%%%%%
%%%%%%%%%%%%%%%%%%%%%%%%%%%%%%%%%%%%%%%%%%%%%%%%%%%%%
%%%%%%%%%%%%%%%%%%%% FIG. 2 %%%%%%%%%%%
%%%%%%%%%%%%%%%%%%%%%%%%%%%%%%%%%%%%%%%%%%%%%%%%%%%%%%%%%%%%

In discussing the phase diagram of Fig.~\ref{Fig:fig2}, we begin from the 
left (weak dipoles) and move to the right (toward stronger dipoles).  The 
formation of vortex-antivortex pairs is energetically unfavorable for weak
dipoles and, hence, $\rho_{pair} = 0$ for the leftmost state.  The next 
configuration corresponds to a single vortex-antivortex pair per unit 
cell with vortices in the vicinity of the dot centers and antivortices centrally
located in the interstitial areas.  With stronger dipoles, there is a 
first order transition to a configuration for which $\rho_{pair} = 2$.
As can be seen in Fig.~\ref{Fig:fig3}(a), antivortices (small dark regions)
are connected with the dot magnets by lobes of depleted Cooper pair 
density (light areas).  These molecule-like complexes are aligned along
the unit cell diagonals, thereby breaking $\pi/2$ orientational symmetry.
As the dipole strengths are increased further, one finds a state which 
breaks both orientational and mirror symmetries [see Fig.~\ref{Fig:fig3}(b)]
.  As the sudden symmetry 
breaking suggests, the transition is first order. 

     Even higher dipole strengths lead to a fractional state, for which
$\rho_{pair} = 2.5$.  Figs.~\ref{Fig:fig3}(b) and~\ref{Fig:fig3}(c) reveal that
the transition into this state is a gradual one.  Ultimately mirror 
symmetry is restored, as seen in Fig.~\ref{Fig:fig3}(c).  As one continues  
the rightward sweep, one encounters a first order transition to a $\rho_{pair} 
= 3$ state [depicted in Fig.~\ref{Fig:fig3}(d)].  Again, in view of the 
significant dissimilarities between the states, this is not surprising.  
For even larger magnetic dot strengths, the vortex pair density exhibits
a surprising nonmonotonicity by jumping suddenly to zero.  Despite 
the abrupt change in $\rho_{pair}$, the shift to the $\rho_{pair} = 0$ state
occurs continuously, through the sequence shown in Fig.~\ref{Fig:fig4}.  
First, the antivortices rearrange to form the $\rho_{pair} = 3$ configuration
depicted in Fig.~\ref{Fig:fig4}(a).  Next, the antivortices are pulled inward,
forming the state shown in Fig.~\ref{Fig:fig4}(b).  The triangular structures in 
Fig.~\ref{Fig:fig4}(b), each of which contains three vortex-antivortex pairs,
then begin to transform.  The vortices move outward from the dots to meet 
the antivortices, which move inward.  Ultimately, the flux quantum pairs
annihilate and the 
result is the $\rho_{pair} = 0$ configuration shown in Fig.~\ref{Fig:fig4}(c).
The final transition, one which also occurs in a continuous manner, is to 
the $\rho_{pair} = 4$ configuration shown in Fig.~\ref{Fig:fig4}(d).  In each 
unit cell four vortex-antivortex pairs form  along nearest neighbor bonds.
As the magnetic dipoles are made
stronger, the pair separations increase, until the magnetic   
dipoles become so strong that the superconductivity in the thin film is 
lost.  

      This complicated evolution is driven by the competition between intra and
interdot potentials for the antivortices.  Evidently, for very strong dipoles,
the antivortices feel not just the effects of a single magnet, but are affected 
by neighboring magnets as well, and the state deforms in such a way that the 
vortices instead reside on nearest neighbor bonds.

%%%%%%%%%%%%%%%%%%%% FIG. 3 %%%%%%%%%%
%%%%%%%%%%%%%%%%%%%%%%%%%%%%%%%%%%%%%%%%%%%%%
%%%%%%%%%%%%%%%%%%% FIG. 4 %%%%%%%%%%%
%%%%%%%%%%%%%%%%%%%%%%%%%%%%%%%%%%%%%%%%%%%%%%

{\it Conclusions and possibilities for experiment--}
We have found, even in the absence of any applied magnetic flux, that 
mesoscopic arrays of magnetic dots can exhibit nontrivial vortex 
phenomena, including configurations which break lattice symmetries, 
states with superlattice structure, and a fractional vortex configuration.  
To realize these states in the laboratory, we have proposed an 
experiment in which one may vary the effective strength of the dot 
magnets by using an in-plane magnetic field to tilt the dipole moment.  
In this way it should be possible to carry out the previously discussed
sweeps in magnet strength, making it feasible to 
access experimentally the stable configurations shown in the phase 
diagram of Fig.~\ref{Fig:fig2} and even some of 
the states which we have classified as metastable. 

Finally, we mention some possible experimental tests for the novel vortex
phenomena discussed in this work.  Critical currents are a useful probe; 
different $j_{c}^{x}$ and $j_{c}^{y}$ values would signal broken orientational
symmetry.  The existence of states that break the spatial symmetry of the 
lattice [e.g., Fig.~\ref{Fig:fig3}(b)] suggests that Ising physics might 
be observed in this system at finite temperatures in thermodynamic 
quantities such as the specific heat.  It is also interesting to speculate 
that the non-monotonic vortex density found in the vicinity of the four
vortex pair states (Figs.~\ref{Fig:fig4}) could lead to a peak in the critical 
current as a function of perpendicular dipole strength.  In any case, our 
calculations strongly suggest that this system offers a rich phenomenology 
worthy of further theoretical and experimental investigation.

{\it Acknowledgements--}
The authors would like to thank M.V. Milo\v{s}evi\'{c}, I. Zuti\'{c}, and 
S. Das Sarma for useful
discussions.  This work was supported by NSF Grant No. DMR01-08451.

\end{document}